\documentclass[twocolumn,superscriptaddress,nobibnotes,nofootinbib]{revtex4-2}
\usepackage[utf8]{inputenc}
\usepackage{amsmath}
\usepackage[english]{babel}
\usepackage[dvipsnames]{xcolor}
\usepackage{amssymb}
\usepackage{textcomp}
\usepackage{graphicx}
\usepackage{placeins}
\usepackage{textgreek}
\usepackage{upgreek}
\usepackage{latexsym}
\usepackage{braket}
\usepackage[nolist]{acronym}
\usepackage{siunitx}
\usepackage{xpatch}
\usepackage{dsfont}
\usepackage{xcolor}
\usepackage{hyperref}
\hypersetup{linktoc=all, hyperindex=true,colorlinks, linkcolor={red!50!black},citecolor={blue!50!black},urlcolor={blue!80!black}}

\makeatletter

\def\l@subsection#1#2{}
\def\l@subsubsection#1#2{}
\patchcmd{\@ssect@ltx}
    {\addcontentsline{toc}{#1}{\protect\numberline{}#8}}
    {}
    {}
    {}
\makeatother

\begin{document}
%
\newacro{lqg}[LQG]{linear quadratic gaussian regulator}
\newacro{lqr}[LQR]{linear quadratic regulator}

\newcommand{\bop}{\ensuremath{b}}
\newcommand{\bdag}{\ensuremath{b^{\dagger}}}
\newcommand{\aop}{\ensuremath{a}}
\newcommand{\adag}{\ensuremath{a^{\dagger}}}
\newcommand{\dd}[1]{\ensuremath{\mathrm{d}#1\,}}
\newcommand{\ii}{\ensuremath{\mathrm{i}}}
\newcommand{\ee}{\ensuremath{\mathrm{e}}}
\newcommand{\ddd}[1]{\ensuremath{\mathrm{d}^3#1\,}}
\newcommand{\mean}[1]{\ensuremath{\langle  #1 \rangle}}
%
\title{Optically levitated gyroscopes with a MHz rotating micro-rotor}

\newcommand{\NUDT}{College of Intelligence Science and Technology, National University of Defense Technology, Changsha, 410073, China}
\newcommand{\wuhanNUDT}{College of Information and Communication, National University of Defense Technology, Wuhan, 430000, China}

%
\author{Kai Zeng}
\affiliation{\NUDT}

\author{Xiangming Xu}
\affiliation{\wuhanNUDT}

\author{Yulie Wu}
\email{ylwu@nudt.edu.cn}
\affiliation{\NUDT}

\author{Xuezhong Wu}
\affiliation{\NUDT}

\author{Dingbang Xiao}
\email{dingbangxiao@nudt.edu.cn}
\affiliation{\NUDT}

\begin{abstract}
The optically levitated particles have been driven to rotate at an ultra-high speed of GHz, and the gyroscopic application of these levitated particles to measure angular motion have long been explored. However, this gyroscope has not been proven either theoretically or experimentally. Here, a rotor gyroscope based on optically levitated high-speed rotating particles is proposed. In vacuum, an ellipsoidal vaterite particle with 3.58 $\mu$m average diameter is driven to rotate at MHz, and the optical axis orientation of the particle is measured by the particle rotational signal. The external inputted angular velocity makes the optical axis deviate from the initial position, which changes the frequency and amplitude of the rotational signal. The inputted angular velocity is hence detected by the rotational signal, and the angular rate bias instability of the prototype is measured to be $0.08^o/s$. It is the smallest rotor gyroscope in the world, and the bias instability can be further improved up to $10^{-9o}/h$ theoretically by cooling the motion and increasing the angular moment of the levitated particle. Our work opens a new application paradigm of the levitated optomechanical systems and possibly bring the rotor gyroscope to the quantum realm.
\end{abstract}

\flushbottom
\maketitle

\thispagestyle{empty}

\section*{Introduction}

The optical levitation system uses the focused laser to trap particles. The maximum light intensity at the focal point generates a gradient force on the particles, directing them towards the center of the focus. This allows for the achievement of stable particle suspension without the need for additional control~\cite{vacuum2021science,tongcang2013doctor,Ashkin1986ol}. If the laser beam is circularly polarized, the angular momentum of the laser can drive the levitated particle to rotate~\cite{quantumrotation2021NRP,liGHz2018prl,Arita2011AC}. By detecting the light passing through the trapped particles, the motion of the trapped particle can be detected~\cite{cavity_cool2015PRL,cavity2015nanoletter,full2017optica}. Therefore, the optical levitation, rotation and detection system can be realized by a single laser beam, which greatly simplifies the device system compared with the liquid levitation~\cite{2012_MEMS_rotor_gyro}, air levitation~\cite{gas2004levitation}, electromagnetic levitation~\cite{2000_rotor_gyro} and electrostatic levitation~\cite{2012_ESG_IEEE}. At the same time, the speed of the levitated particles can be driven to GHz level by reducing the ambient air pressure~\cite{li2020NT,GHz2021PR,GHz2018PRL_ETH}. Based on the optical levitation system~\cite{2020_review_Optomechanics}, high-precision measurement of force~\cite{force2016PRA}, torque~\cite{li2020NT} and acceleration~\cite{accelaration2017PRA} has been realized, and as the thermal motion of the trapped particles is cooled to the quantum ground state, the measurement accuracy is expected to be further improved~\cite{2020_science_quan_cool,quancontrol2021nature,quancool2021nature1}. 

High-speed rotating objects exhibit a stable spin axis in the space, which can be used to measure angular motion. This phenomenon is know as the gyroscopic effect. This effect is of great significance in engineering and navigation. Compared with vibration gyroscope (gyro) and optical gyro, the rotor gyro has the longest history of development. However, its performance and accuracy have consistently remained at a leading level. In the case of optically levitated rotating particles, the gyroscopic stabilization of translational motion has been observed and exploited in the MHz rotation of vaterite micro-sphere~\cite{Arita2013NC} and the GHz rotation of silica nano-dumbbells~\cite{liGHz2018prl}. This trait of rotating particle can be used to cool their rotations to sub-Kelvin temperatures~\cite{five2020li,cool2021prl} and frequency-lock the rotation to an external drive so as to obtain ultra-stable nanomechanical rotor~\cite{ultrastable2017NC}. Since the trapping laser tends to align the particle major axis with the laser electric field~\cite{xiguang2016JOSA}, the precession of rotating objects is also investigated with nano-scale silica particles~\cite{precession2018PRL,five2020li}. Furthermore, the novel strategies for precision tuning of the orientation of levitated microspheres are proposed~\cite{2021_xie_SA}, which is indispensable for the realization of micromotors and microgyroscopes. Despite extensive research on the dynamics of optically levitated rotating rotors, such as gyroscopic stabilization and precession, there is a lack of theoretical and experimental studies on the measurement of angular motion in the presence of specific obstacles. Firstly, the optical torque that aligns the particle optical axis with the laser electric field~\cite{xiguang2016JOSA,precession2018PRL} makes the rotor precessed, which is different from the traditional free mechanical rotor gyroscope. In addition, the measurement of spin axis of optically levitated particles is challenging for its tiny volume. Here, by overcoming these obstacles, we successfully demonstrated the practical application of optical levitation technology in the measurement of angular motion. This represents the first reported implementation of an optically levitated gyroscopes with a rotating micro-rotor.

In this paper, the gyroscopic dynamics of the optically levitated rotor are analyzed, and the influence of rotor’s optical axis and geometric long axis on the attitude angle is investigated. Then, based on the precession of high-speed rotor, the gyroscope response is deduced when there is an external angular velocity. According to the theoretical model, an optically levitation system is built in vacuum to realize the ultra-fast rotation of the trapped particle, and the gyroscopic effect is validated by detecting the spin axis according to the rotational signal. Although the experiments are implemented in the free space optical system, the presented optically levitated gyroscope (OLG) has the potential to realize on a chip~\cite{light2021metafiber,earth2020NP,chip2021optica_li,chip2021optica_yu}.

\section*{Results}
\subsection*{Optical axis alignment}
\begin{figure}
    \centering
    \includegraphics[scale=0.9]{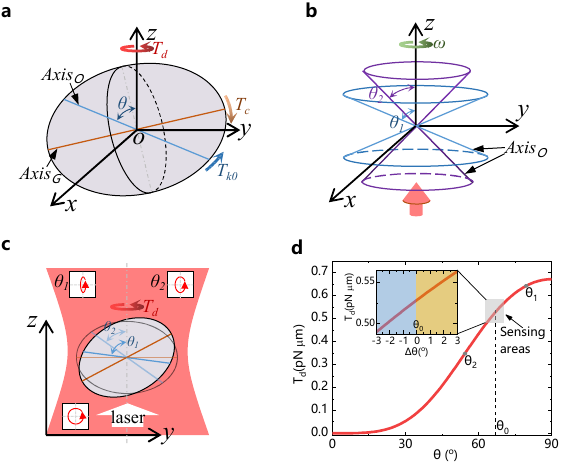}
    \caption{\textbf{Optical axis alignment of an ellipsoidal vaterite particle.} \textbf{a}, An ellipsoidal birefringent particle with optical axis $Axis_O$ and geometrical long axis $Axis_G$.The centrifugal torque $T_c$ comes from the rotation of particle, and balanced with the restoring torque $T_{k0}$. \textbf{b}, Trajectory of optical axis under rotation. Due to the rotation, the trajectory of the optical axis is cone-shaped. The color represents the different angle $\theta_1$ and $\theta_2$. \textbf{c}, The sketch map of laser polarization before and after transmitting the particle. The red ring with arrow inside box represents the polarization ellipse of laser, the lower left circle represents the incident light is circular polarized, and the upper two ellipse represents the polarization of transmitted laser with the angle $\theta_1$ and $\theta_2$. \textbf{d}, Driving torque $T_d$ versus angle $\theta$. The sensing areas means the gyroscope application analyzed below. The simulation parameters are presented in supplementary note \ref{sec:note2}.}
    \label{fig:1}
\end{figure}

As shown in Fig. \ref{fig:1}a, assuming the laser propagates along the z-axis in the positive direction, the optical torque exerted by the circularly polarized laser comprises two components~\cite{xiguang2016JOSA,precession2018PRL,2021_xie_SA}. On the one hand, a driving torque $T_d$ drives the particle to spin, and the spin frequency is determined by the balance between the friction torque from the surrounding air and the driving torque from laser beam. On the other hand, a restoring torque $T_{k0}$ aligns the optical axis $Axis_O$ of the ellipsoidal vaterite particle parallel to the transverse plane (the plane of rotating electric field, $x$-$o$-$y$ in Fig. \ref{fig:1}a) of laser. Therefore, the angle $\theta$ between $Axis_O$ and the laser propagation direction ($z$ axis) is near $90^o$ when the particle dose not rotate. The blue line $Axis_O$ in Fig. \ref{fig:1}a represents the optical axis of the particle with birefringent material, and the brown line $Axis_G$ represents the geometric long axis. Due to the non-ideal spherical shape of particles and the relationship between the optical axis and the growth direction of birefringent crystals, the optical axis and the geometric long axis of the particle typically do not align.

As the spin frequency of the particle increases gradually, due to the geometric asymmetry, the centrifugal torque $T_c$ (makes the heavier end to the rotation plane) will increase, making the optical axis deviate from the transverse plane~\cite{2021_xie_SA}. When the spin frequency is stable, the centrifugal torque $T_c$ and the restoring torque$T_{k0}$ are balanced, and the angle $\theta$ is maintained at a specific position $\theta_0$$\neq$$90^o$. The restoring torque~\cite{xiguang2016JOSA} is $T_{k0}\propto\sin(2\theta)E^2$, where $E$ represents the laser electric field intensity. When the particle oscillates slightly near the equilibrium position, the restoring torque can be approximated as linear $T_k\approx k\Delta\theta$. Analogous to the torsion spring with a stiffness coefficient $k$, and the orientation of the optical axis is limited to the vicinity of its initial position. The stiffness coefficient $k$ is related to the particle shape, material, laser power and so on. Although the optical axis rotate with the particle rotation, the angle $\theta$ keeps unchanged, which can be seen in Fig. \ref{fig:1}b.

As shown in Fig. \ref{fig:1}c, the ellipticity of the polarization ellipse represents the polarization of laser. During the deviation of $\theta$, the polarization of the transmitted light is changed. Since the driving torque $T_d$ comes from the transfer of spin angular moment of laser, the changes of polarization results in the variation of $T_d$. The driving torque is expressed as~\cite{2021_xie_SA}
\begin{equation}
T_d=c_1\sin\left(c_2\left(1-\frac{n_e}{\sqrt{n_o^2\sin^2\theta+n_e^2\cos^2\theta}}\right)\right)
\label{driveT}
\end{equation}

\noindent
where $c_1$ and $c_2$ are constant parameters that related to the laser and particle parameters, $n_e$ and $n_o$ are the refractive indices of the birefringent vaterite particle (see supplementary note \ref{sec:note2} for details). As shown in Fig. \ref{fig:1}d, when the angle $\theta$$=$$90^o$, the driving torque $T_d$ reaches its maximum value. As the spin frequency increases, the optical axis deviates from the transverse plane, and $T_d$ decreases. According to the relation between $T_d$ and $\theta$, the orientation $\theta$ of optical axis can be obtained by monitoring the driving torque, while the driving torque can be measured by the spin rotational signal of particles~\cite{2021_xie_SA,torque2001measure}.

\subsection*{Gyroscopic dynamics}
\begin{figure*}
    \centering
    \includegraphics[scale=1]{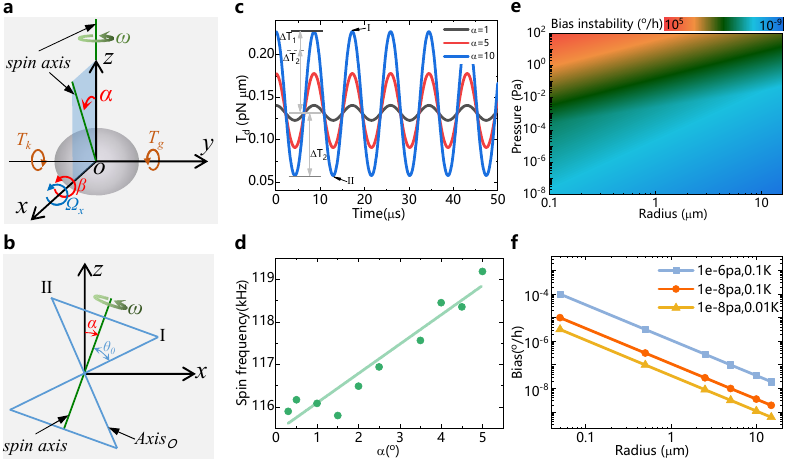}
    \caption{\textbf{Principle of optically levitated rotor gyroscope.} \textbf{a}, Coordinates definition of the gyro. $\alpha$ and $\beta$ are angular displacements, $T_k$ is the restoring torque and balanced with gyroscopic torque $T_g$ that comes from the precession. The external rotation $\Omega_x$ makes the spin axis deflected with $\alpha$, which further changes the angle $\theta$. The alteration of $\theta$ results in the variation of $T_d$, and the $T_d$ can be measured by the particle rotational signal. By disregarding the intermediate steps, the measurement of the external rotation $\Omega_x$ is accomplished through the analysis of the rotational signal. \textbf{b}, Side view of spin axis with a deflection angle $\alpha$. The cone represents the rotational trajectory of the optical axis, and the angle $\theta$=$\theta_0$+$\alpha$ in position \uppercase\expandafter{\romannumeral1}, and $\theta$=$\theta_0$-$\alpha$ in position \uppercase\expandafter{\romannumeral2}. \textbf{c}, Driving torque $T_d$ versus time under different deflection angle $\alpha$. $T_d$ reaches maximum and minimum value in position \uppercase\expandafter{\romannumeral1} and \uppercase\expandafter{\romannumeral2} respectively. The average torque increases with $\alpha$, and the fact that $\Delta$$T_1$ is lager than $\Delta$$T_2$ shows the increase of average torque. The initial angle $\theta_0$$=$$45^o$ and the initial spin frequency is 116kHz in the simulation. \textbf{d}, Spin frequency versus deflection angle $\alpha$. The nonlinear error comes from the numerical solution of differential equations (\ref{Equ_omega}). \textbf{e-f}, Calculated bias instability of optically levitated rotors with different parameters (see the Supplementary note \ref{sec:note3}).}
    \label{fig:2}
\end{figure*}

Different from the free rotor gyroscope, the optically levitated particle is subjected to the restoring torque $T_k$. Except the restoring torque $T_{k0}$ that balanced with the centrifugal torque, the rotor is sustained additional restoring torque $T_k$ when the rotor precessed with the inputted angular motion. The dynamical model, working principle and theoretical bias instability of the optically levitated rotor gyro is presented in Fig. \ref{fig:2}.

Assuming the spin frequency is $\omega$, the moment of inertia is $I$, and the angular momentum of the particle is $H$$=$$I\omega$. For an ideal circularly polarized Gaussian beam, the restoring torque stiffness coefficient along the $x$ and $y$ axes are equal $k_x$$=$$k_y$$=$$k$. The influence of sphericity error of particles on their rotational damping is also negligible, so the damping coefficient can also be approximately equal $c_x$$=$$c_y$$=$$c$. The rotational angles of the particles along the $x$ and $y$ axes are $\beta$ and $\alpha$, and $T_{th}$ is the thermal fluctuation torque. When the angular velocity $\Omega_x$ is applied along the $x$-axis, the gyroscopic dynamic equation of the particle is established as~\cite{lihai2016twodof,twodofgyro2018poletkin,twodofgyro2015poletkin}

\begin{equation}
\begin{cases} 
H\dot{\alpha}-c_x\dot{\beta}-k_x\beta=T_{th}\\
H\dot{\beta}+c_y\dot{\alpha}+k_y\alpha=H\Omega_x+T_{th}
\end{cases}
\label{motion}
\end{equation}

\noindent
where $H\Omega_x$ represents the gyroscopic (precession) torque $T_g$ along the $y$-axis. If the thermal fluctuation torque is ignored, the solution to an input angular velocity $\Omega_x$ can be obtained (Supplementary note \ref{sec:note1}). The rotation angle $\beta$ is transient, while a steady deflection angle $\alpha$$=$$\Omega_xH/k$ will be generated in the orthogonal direction, and it is linear with the input angular velocity. Therefore, as shown in Fig. \ref{fig:2}b, the input angular velocity can be detected by measuring the deflection angle $\alpha$ of spin axis. The angle between the optical axis and the spin axis keeps unchanged as $\theta_0$, but the angle $\theta$ between the optical axis and the laser propagation direction changes with $\alpha$. Assuming the initial position of optical axis is coincide with the $x$ axis (position \uppercase\expandafter{\romannumeral1}), the angle $\theta$ meets the following equation

\begin{equation}
\cos\theta=\cos\theta_0\cos\alpha-\sin\theta_0\sin\alpha\cos(\omega t)
\label{Equ_theta}
\end{equation}

Hence, the driving torque $T_d$ fluctuates as shown in Fig. \ref{fig:2}c due to the changes of $\theta$. The average driving torque increases with $\alpha$, which results in the variation of spin frequency. The rotational dynamics equation of spin is 

\begin{equation}
\frac{d\omega}{dt}I=T_d-c\omega 
\label{Equ_omega}
\end{equation}

\noindent
Substitute equation (\ref{driveT}) and (\ref{Equ_theta}) into equation (\ref{Equ_omega}), the spin frequency under different deflection angle $\alpha$ is shown in Fig. \ref{fig:2}d. Since $\alpha$ is related to the input angular velocity $\Omega_x$, the spin frequency is hence regarded as the output of the gyroscope in the following experiment. It is worth noting that the changes of the driving torque are related to the initial angle $\theta_0$, hence the sensitivity can be modulated by controlling the initial stable angle.

As shown in Fig. \ref{fig:2}e-f, taking the thermal fluctuation torque into account in the dynamic equation (\ref{motion}), the bias instability of this gyroscope can be estimated by the equivalent angular velocity $\Omega_{th}$$=$$T_{th}/H$ (Supplementary note \ref{sec:note3}). The bias instability is $B$$=$$k_{\sigma}\Omega_{th}$$\propto$$(T_{emp}P_0/a^3)^{1/2}$, where $k_{\sigma}$ is the bias instability conversion coefficient, $T_{emp}$ is the torsional vibration equivalent temperature, $P_0$ is the pressure, and $a$ is the radius of the rotor. Therefore, the bias instability can be improved by decreasing the motion temperature~\cite{five2020li} and pressure or increasing the particle size, and the instability of $10^{-9o}/h$ is possible for a 30$\mu$m diameter particle with the torsional vibration equivalent temperature of 0.1$K$ under $10^{-8}$pa.

\subsection*{Experimental verification}
\begin{figure*}
    \centering
    \includegraphics[scale=0.8]{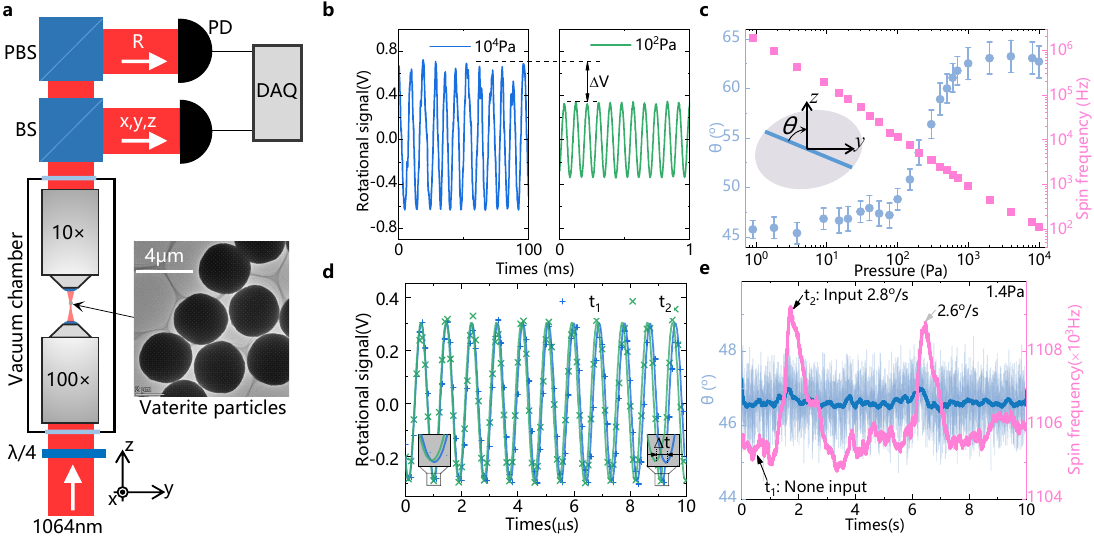}
    \caption{\textbf{Experimental setup and the measurement results of alignment and gyro effects.} \textbf{a}, Sketch map of experimental setup. Labels denote the quarter wave plate ($\lambda$/4), microscope object (100× and 10×), beam splitter (BS), polarizing beam splitter (PBS), photo detector (PD), data acquisition card (DAQ). The inset are vaterite particles used as rotor. \textbf{b}, Spin rotational signal of trapped particle at different pressure, and the amplitude is related to $\theta$. \textbf{c}, $\theta$ and spin frequency versus pressure. $\theta$ decreases with the decrease of pressure, which means the optical axis rotates towards the propagation direction of laser. \textbf{d}, Spin rotational signal at different moment $t_1$ and $t_2$, and the angular velocity is inputted at moment $t_2$. \textbf{e-f}, Changes of $\theta$ and spin frequency under inputted angular velocity.}
    \label{fig:3}
\end{figure*}

An optical levitation system (Fig. \ref{fig:3}a, see Methods section for details) is established to verify the proposed theoretical model. The vaterite particle with a mean diameter of 3.58$\mu$m and a standard deviation of 0.19$\mu$m is trapped in vacuum and driven to spin with frequency of 100-1000$kHz$. The spin and center of mass motion are measured by photo detector and recorded by data acquisition card. In order to input angular velocity and isolate the vibration of ground, the setup is built on an anti-vibration air-floating platform.

As mentioned above, the initial positon $\theta_0$ will change under the action of centrifugal torque during the process of spin rotation acceleration. Fig. \ref{fig:3}b shows two spin rotational signals of a particle at pressures of $10^4P_a$ and $10^2P_a$. The spin frequency increases with the the decrease of pressure, while the amplitude of the rotational signal decreases. According to the optical torque measurement method~\cite{2021_xie_SA,torque2001measure}, the lager amplitude represents the lager torque. Meanwhile, the angle $\theta$ is related to the torque as shown in Fig. \ref{fig:1}d, hence the decreases of amplitude indicates a corresponding change in $\theta$. Since the spin frequency is also sensitive to the pressure, the angle $\theta$ is measured by the amplitude when decreasing the pressure. Fig. \ref{fig:3}c is the measured $\theta$ under different pressure, and the transform method from amplitude to $\theta$ is presented in Supplementary note \ref{sec:note2}. With the decrease of air pressure, the spin frequency increases gradually, resulting in the increase of centrifugal torque, so the optical axis of particles gradually moves away from the initial position. When the pressure below $100P_a$, the polar angle tend to stable around $46^o$.

When the pressure remains stable, the spin frequency is utilized to represent the deflection angle $\alpha$ induced by the input angular velocity. Fig. \ref{fig:3}d is the comparison of spin rotational signal with ($t_2$) and without ($t_1$) the external inputted angular velocity. The amplitude of the two signal is approximately  equal cause the deflection angle is too small, while the periods exhibit inequality, which means the frequency is more sensitive to the external angular velocity. The time difference $\delta{t}$ proves the changes of frequency. As shown in Fig. \ref{fig:3}e, the frequency and amplitude of the spin rotational signal are obtained by fitting (see Methods), and the frequency changes about 4$kHz$ when there is an external angular velocity $2.8^o/s$. Since the deflection angle is about $0.5^o$, the variety of amplitude is about 8 mV (smoothed line). However, the standard deviation of amplitude is about 10mV, which is beyond the variety.

\subsection*{Gyroscope performance}
\begin{figure}
    \centering
    \includegraphics[scale=0.7]{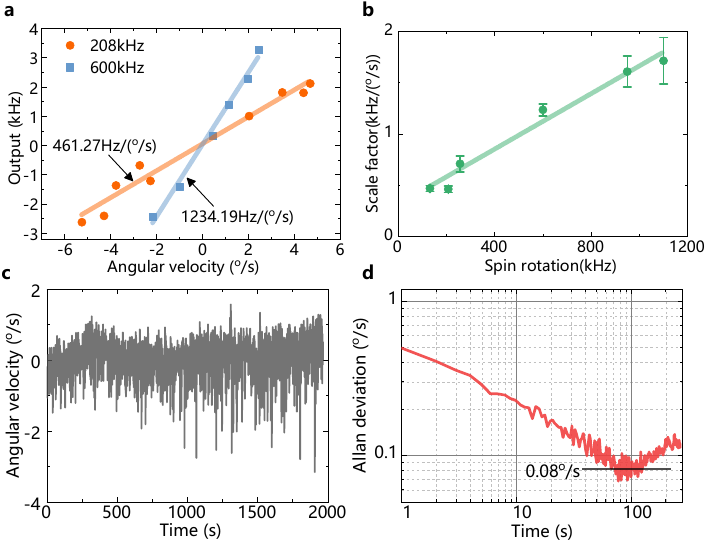}
    \caption{\textbf{Gyroscope performance of the OLG prototype.} \textbf{a}, The output of OLG under different external angular velocity. Extended data are presented in supplementary Fig. ~\ref{fig:S7}. \textbf{b}, Scale factor under different spin frequency. The error bars are determined by fitting. \textbf{c}, Measured angular velocity with no input, which shows the instability of OLG. The spin frequency of the rotor is 470$kHz$ at 4$Pa$, and the sampling rate is 1$Hz$. \textbf{d}, Allan deviation of the prototype, and the bias instability is $0.08^o/s$.}
    \label{fig:4}
\end{figure}

The gyroscopic effect is evaluated under different stable spin frequencies (stable pressure). As shown in Fig. \ref{fig:4}a, the changes of spin frequency roughly corresponds linearly to the input angular velocity. There are two main reasons for the inconsistency in the measurement. One is the transient response of angle to the input angular velocity, and the other is the fluctuation of spin frequency caused by the changes of ambient air pressure. As shown in Fig. \ref{fig:4}b, the response at different spin frequency shows that the scale factor of the optically levitated gyroscope increases with the increase of spin frequency, which is consistent with the theoretical results. Due to the deflection angle of the gyroscope is $\alpha$$=$$\Omega{H}/k$, when the spin frequency of particles increases, the angular momentum $H$ will increase, resulting in the increase of sensitivity $H/k$. For example, the scale factor is $461.27Hz/(^o/s)$ when the spin frequency is 208$kHz$, while the scale factor is $1234.19Hz/(^o/s)$ when the spin frequency increase to 600$kHz$. The difference between the ratios of scale factor (2.68) and spin frequency (2.88) is about $7\%$.

In addition, we also tested the bias instability of the optically levitated rotor gyro. The output signal without the external angular velocity during half an hour is presented in Fig.~\ref{fig:4}c. It can be seen that the output signal (i.e. the product of spin frequency variation and scale factor) has high-frequency fluctuation and long-period slow change. The fluctuation around the mean value is mainly caused by the thermal motion of particles, while the slow change is the result of the changes of pressure and laser power during the test. Furthermore, the output signal mainly dropped sharply, but did not rise sharply. For example, the output fluctuation reached $-3^o/s$ at 1800 seconds. This is because the thermal motion of the particle causes them to deviate from the laser focus, which is equivalent to the reduction of the laser power, and which reduces the spin frequency and ultimately affects the stability of the test~\cite{frequency_stable2020PRA}. As shown in Fig. \ref{fig:4}d, the Allan variance curve shows that the bias instability of the prototype is about $0.08^o/s$. By improving the above factors, the stability of the optically levitated gyro can be improved~\cite{ultrastable2017NC,beamcharac2008AO}.

\section*{Discussion}
The first operation and evaluation of a micro rotor gyroscope based on optically levitated particles is presented. The working principle of this gyro is proposed theoretically according to the response of the spin axis to the input angular motion, and the gyroscopic effect is verified experimentally according to the measurement of spin axis using the rotational signal. These results confirm the possibility of optically levitated rotor gyroscope.
Futhermore, this paper has developed the dynamics and sensing applications of the levitated optomechanical system, and it also brings new development opportunities for the traditional mechanical rotor gyroscope. At the same time, the acceleration measurement based on levitated particles has developed relatively mature~\cite{pu2021micro,force_accelaration_sensing2020PRA}. Combined with the angular velocity measurement scheme proposed here, a single levitated particle can complete the functions of accelerometer and gyroscope to achieve inertial navigation. Compared with multi-axis combined inertial sensing technology, this scheme can greatly simplify the complexity of the system and is expected to become a new generation of high-precision measuring elements. Although the sensitivity and instability of the gyroscope are not ideal at present, the performance of the gyroscope can be further greatly improved by laser cooling, customized levitated rotor and enhanced optical trap stiffness~\cite{kuang2023NP,miao2022OE,zhengyu2020PRL}, which providing new sensing technology for the inertial navigation field.

\section*{Methods}
\subsection*{Experimental setup}
As shown in Fig.~\ref{fig:S5}a and b, the optically levitated rotor gyroscope (OLG) and a commercial gyroscope (InvenSense ICM-20602) are both installed on the platform. The experimental optical path is presented in Fig.~\ref{fig:S5}a, and the red and green line represents the trapping and imaging light separately. The vertical optical levitation system is utilized in the experiment. The trapping laser (LASEVER, LSR1064NL, wavelength 1064nm, power 300$mW$) is focused by a high numerical aperture microscope object (Nikon, E Plan ×100, NA=1.25) in the vacuum chamber. The laser power used in the experiment is about 20mW. Before the laser enters the vacuum chamber, the beam expander is used to expand the beam diameter to about 1.5 times of the pupil behind the objective lens, so as to make full use of the focusing ability of the objective lens and improve the trap stiffness. Then the laser power is adjusted by the combination of 1/2 wave plate and polarizing beam splitter (PBS1), and PBS1 also plays the role of adjusting the laser polarization. The extinction ratio of linearly polarized light from the PBS is TP:TS$>$3000:1. This works together with the subsequent 1/4 wave plate to ensure a high degree of circular polarization of the laser entering the vacuum chamber.

The laser is focused by the high numerical aperture objective in the vacuum chamber, and particles are trapped near the focus. The vibration and rotation of the trapped particles will modulate the laser passing through the particles. The transmitted light is collected by a low-NA objective lens and reflected by the dichroic mirror to the detection part. In this case, the transmitted light is divided into four beams to detect the particle rotation signal ($R$) and three-dimensional vibration signal ($xyz$). The polar angle of the particle optical axis is read out by the amplitude or frequency of the rotational signal. The rotational signal (polarization modulated) of particles is distinguished after passing through PBS2, so that the signal of rotation motion can be read out by detecting its power changes with photo detector. The vibration signal of the $xyz$ axes are obtained by differential ways. The movement in the $x$ and $y$ directions are distinguished by D-shaped mirrors in different directions, and the movement in the z direction changes the focus position of the beam by the refraction of particles, resulting in the change of light intensity on the probe. As shown in Fig.~\ref{fig:S5}c, the resonant frequency of the trapped particle is about $f_x$=430$Hz$, $f_y$=390$Hz$ and $f_z$=200$Hz$.

\subsection*{Sample preparation}
The birefringent material of particles used in the experiment is vaterite, which is synthesized by solution method and has a mean diameter of 3.58$\mu$m and a standard deviation of 0.19$\mu$m. The particle shape and elemental mapping are shown in Fig.~\ref{fig:S6}. In the experiment, the particles are placed in a small chamber (without sealing, which ensures ventilation and reduces the impact of airflow disturbance on particles), and the small chamber is installed on the piezoelectric plate. Through the vibration of piezoelectric plate, the particles are thrown near the laser focus to achieve trapping.

\subsection*{Gyroscopic effect experiments}
The comparison results of the optically levitated gyro (OLG) and the commercial gyro are shown in Fig.~\ref{fig:S7}. The output of the OLG is the changes of the spin frequency of the levitated rotor. The spin frequency is obtained by fitting the rotational signal as shown in Fig.~\ref{fig:S8}a, and it is fitted every 1 $ms$. Since the frequency resolution of Fourier Transform is 1/T, T is the sampling time, the frequency space is 1$kHz$ for 1$ms$. Nevertheless, the frequency resolution of fitting is more precise. As shown in Fig.~\ref{fig:S8}b, we use the signal generator to test the resolution of fitting method. The initial output frequency is 100$kHz$, and then the frequency is changed step by step (10$mHz$). It is clear that the frequency resolution of the fitting method is below 10$mHz$.

\section*{Acknowledgements}
This project was supported by the National Natural Science Foundation of China (51975579).

\section*{Author contributions} K.~Z. and Y.~W. designed and built the experiment. K.~Z. and D.~X. performed the measurements. K.~Z., X.~X., and X.~W. analyzed the data and all authors contributed to writing and editing of the paper. 

\section*{Data Availability} The data that support the plots within this paper and other findings of this study are available from the corresponding author upon reasonable request.

\section*{Competing Interests} The authors declare no competing financial interests.

\let\oldaddcontentsline\addcontentsline
\renewcommand{\addcontentsline}[3]{}
\bibliography{refs}
\let\addcontentsline\oldaddcontentsline


\clearpage
\onecolumngrid

\renewcommand{\thesection}{S\arabic{section}}  
\renewcommand{\thetable}{S\arabic{table}}  
\renewcommand{\thefigure}{S\arabic{figure}} 
\renewcommand{\theequation}{S\arabic{equation}} 
\setcounter{figure}{0}
\setcounter{equation}{0}
\setcounter{section}{0}
\setcounter{table}{0}
\hypersetup{colorlinks,linkcolor={red!50!black},citecolor={blue!50!black},urlcolor={blue!80!black}}

\begin{center}
\large\textbf{\uppercase{Supplementary Information}}
\end{center}

\section{Supplementary Note 1: Theoretical model}
\label{sec:note1}

\begin{figure}[!h]
\includegraphics[scale=1]{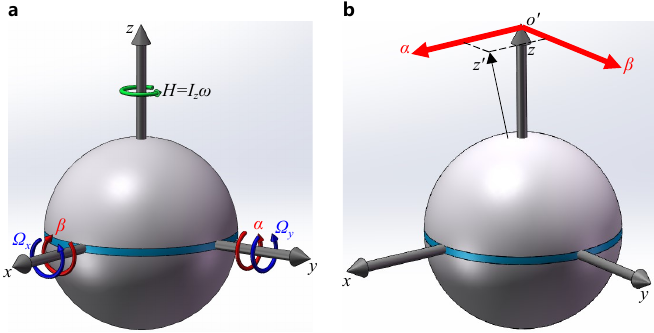}
\caption{\textbf{Definition of Coordinate system.}.\textbf{a}, Sketch map of rotational angle $\alpha$ and $\beta$, external angular velocity $\Omega_x$ and $\Omega_y$, angular moment $H$. \textbf{b}, Phase plane representation of particle rotation. The blue part is used for vision.}
\label{fig:S1}
\end{figure}

Fig. \ref{fig:S1}a shows the definition of the coordinate system of optically levitated particles, the coordinate origin $O$ is the center of particles, and the $xyz$ axes are the radial direction of particles. In the following, $I_j$ ($j$$=$$x,y,z$) represents the moment of inertia, $k_j$ represents the optical torsion spring stiffness~\cite{xiguang2016JOSA,precession2018PRL,2021_xie_SA}, $c_j$ represents the particle rotational damping coefficient, and $\Omega_j$ represents the input angular velocity, $\alpha$ and $\beta$ represents the rotational angle of particles along the $y$ and $x$ axes, respectively. The positive rotation direction of $\alpha$ is consistent with the $y$-axis, but $\beta$ is opposite to the $x$-axis. As shown in Fig. \ref{fig:S1}b, the direction of phase plane $\alpha$-$o'$-$\beta$ can be consistent with the direction of $x,y$ with the above definition of $\alpha$ and $\beta$. It is assumed that the circularly polarized laser propagates along the positive direction of the $z$ axis, and the particle rotates around the $z$ axis under the driving torque with angular frequency $\omega$ , the angular moment of particles is $H=I_z\omega$.

\begin{figure}[!h]
\includegraphics[scale=1]{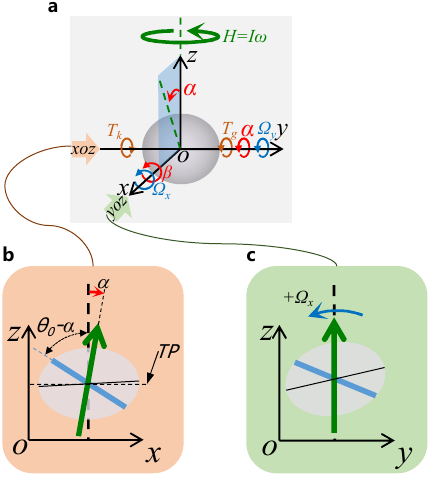}
\caption{\textbf{Sketch map of the gyroscope working principles.}.\textbf{a}, Theoretical model for dynamic analysis. When there is only an angular velocity $\Omega_x$ in $x$ axis, the particle would follow the rotation in the direction of input angular velocity as \textbf{c}, while the orthogonal direction would produce a deviation $\alpha$ as \textbf{b}. $TP$ represents the transverse plane of laser.}
\label{fig:S2}
\end{figure}

The theoretical model is presented as shown in Fig. \ref{fig:S2}. Since this paper mainly discusses the rotational motion of the particle spin axis, only the torque acts on the particle is analyzed. In the direction of $x$-axis, the particle is subjected to the thermal noise torque $T_{th}$, and the torque corresponding to the angular acceleration $-I_x\dot{\Omega}_x$, $I_x\ddot{\beta}$, gyroscopic torque $-H\Omega_y$, $-H\dot{\alpha}$, damping torque $c_x\dot{\beta}$, restoring torque $k_x\beta$.The torque in the $y$-axis is similar to that, so the dynamic equation is established~\cite{lihai2016twodof,twodofgyro2015poletkin,twodofgyro2018poletkin}

\begin{equation}
\begin{cases}T_{th}-I_x\dot{\boldsymbol{\Omega}}_x+I_x\ddot{\boldsymbol{\beta}}-H\boldsymbol{\Omega}_y-H\dot{\boldsymbol{\alpha}}+\boldsymbol{c}_x\dot{\boldsymbol{\beta}}+k_x\boldsymbol{\beta}=0\\T_{th}-I_y\dot{\boldsymbol{\Omega}}_y-I_y\ddot{\boldsymbol{\alpha}}+H\boldsymbol{\Omega}_x-H\dot{\boldsymbol{\beta}}-\boldsymbol{c}_y\dot{\boldsymbol{\alpha}}-k_y\boldsymbol{\alpha}=0&\end{cases}
\label{Equ_S1}
\end{equation}

For an ideal circularly polarized Gaussian beam, the stiffness coefficients of particles in different directions are equal $k_x=k_y=k$. The influence of sphericity error of particles on their rotational damping is also slight, so the damping coefficient can also be approximately equal to $c_x=c_y=c$. Assuming that the input angular velocity is a constant value, the torque corresponding to the angular acceleration $-I_x\dot{\Omega}_x$ can be ignored. The torque $I_x\ddot{\beta}$ corresponding to the nutation of the particle spin axis can also be ignored because its frequency is twice the rotation and its amplitude is small. It is further simplified as that there is angular velocity $\Omega_x$ input in the $x$-axis direction only, and the thermal noise torque is ignored. The simplified motion equation of the optically levitated rotor is

\begin{equation}
\begin{cases}H\dot{\alpha}-c_x\dot{\beta}-k_x\beta=0\\H\dot{\beta}+c_y\dot{\alpha}+k_y\alpha=H\Omega_x\end{cases}
\label{Equ_S2}
\end{equation}

The equation is solved as

\begin{equation}
\begin{cases}\alpha=&\frac{H}{k}\Omega_x-\frac{H}{k}\Omega_x\exp\!\left(\frac{-ck}{H^2+c^2}t\right)\cos\!\left(\frac{kH}{H^2+c^2}t\right)\\\beta=&\frac{H}{k}\Omega_x\exp\!\left(\frac{-ck}{H^2+c^2}t\right)\sin\!\left(\frac{kH}{H^2+c^2}t\right)\end{cases}
\label{Equ_S3}
\end{equation}

It can be seen that when there is an external angular velocity in the $x$-axis, the spin axis of particles will respond around the $x$-axis and $y$-axis simultaneously. But in the same direction, the angle $\beta$ (rotation about the $x$-axis) is an oscillating rotation (libration) with attenuation. Although there is also an oscillation term about $\alpha$ along the $y$-axis, it will eventually produce a stable deflection angle $\Omega_x{H/k}$. In other words, when the angular velocity $+\Omega_x$ is input along the positive direction of $x$ axis (Fig. \ref{fig:S2}c), the particle would rotate in the positive direction of $y$ axis (Fig. \ref{fig:S2}b).

\begin{figure}[!h]
\includegraphics[scale=1]{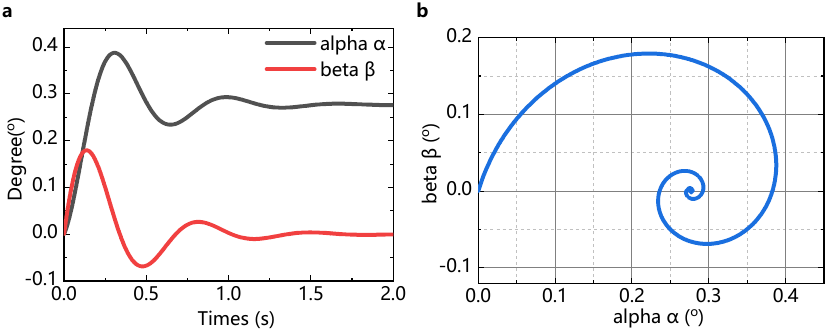}
\caption{\textbf{Simulation results of the rotational angle.}.\textbf{a}, Rotation response to the input angular velocity $\Omega_x$. \textbf{b}, Motion track of particle spin axis endpoint. The spin rotation frequency is 1$MHz$, and the rotor is a 4$\mu$m diameter vaterite spherical particle.}
\label{fig:S3}
\end{figure}

The specific change rule is shown in Fig. \ref{fig:S3}a, and the angular velocity of 2 $^o/s$ is input along the positive direction of the $x$-axis at zero time. It can be found that the particle rotate along the $x$ and $y$ axes simultaneously, but the rotation along the $x$ axis gradually tends to 0, and the rotation along the $y$ axis gradually stabilizes at a fixed angle. The motion track of the spin axis endpoint is shown in Fig. \ref{fig:S3}b, it can be seen that the spin axis finally stabilized in a spiral motion.The spiral motion of the spin axis is mainly caused by the restoring torque and damping torque. As shown in formula (\ref{Equ_S3}), the angular frequency of spiral rotation is $kH/(H^2+c^2)$, and the attenuation constant of rotation amplitude is $ck/(H^2+c^2)$.

From the above analysis, it can be found that after inputting angular velocity in a certain direction, a stable angle will be generated in its orthogonal direction. Due to the existence of the angle, the rotor is subjected to a restoring torque, which makes the rotor generate the precession following the input angular velocity, thus reaching a stable state. In this state, by detecting the deflection angle $\alpha=\Omega_xH/k$ in the orthogonal direction, the input angular velocity $\Omega_x$ can be obtained, thus realizing the measurement of external rotation by optically levitated particles.

\section{Supplementary Note 2: Optical axis measurement}
\label{sec:note2}

The measurement of the deflection angle is realized by detecting the rotational signal under different driving torque as shown in Fig. \ref{fig:1}d. The driving torque $T_d$ is~\cite{2021_xie_SA}

\begin{equation}
\label{Equ_S4}
T_d=\frac{PB_0}{\nu}\hat{z}\sin^2\Bigg(\frac{2\pi an_o}{\lambda}\Bigg(1-\frac{n_e}{\sqrt{n_0^2\sin^2\theta+n_e^2\cos^2\theta}}\Bigg)\Bigg)
\end{equation}

\noindent
where $P$ is the laser power, $B_0$ is the overlap between the trapping beam and the particle, $a$ is the particle radius, $\nu$ and $\lambda$ are frequency and wavelength of the laser separately, $n_e$ and $n_o$ are the refractive indices for laser polarized along the extraordinary and ordinary axes. The simulation results in Fig. \ref{fig:1}d in the main text used $P$=20$mW$, $B_0$=0.0785, $\lambda$=1064$nm$, $n_e$=1.65 and $n_o$=1.55.

The rotational signal is~\cite{2021_xie_SA}

\begin{equation}
\label{Equ_amplitude}
V=\frac{T_d\boldsymbol{\upsilon}G}4\cos(2\omega t)
\end{equation}

\noindent
where $G$=3$kV/W$ is the conversion gain of the photo detector in our experiment. According to the measured amplitude of the rotational signal and the experimental parameters mentioned above, the angle $\theta$ can be obtained as shown in Fig. \ref{fig:3}.

\section{Supplementary Note 3: Bias instability analysis}
\label{sec:note3}

When there is no angular velocity, but only the thermal noise torque is considered, the dynamic equation of the rotating particle can be expressed as

\begin{equation}
\label{Equ_S5}
\begin{cases}H\dot{\alpha}-c_x\dot{\beta}-k_x\beta=T_{th}\\H\dot{\beta}+c_y\dot{\alpha}+k_y\alpha=T_{th}&\end{cases}
\end{equation}

It can also be solved that the steady-state solution is

\begin{equation}
\label{Equ_S6}
\begin{cases}\alpha_{th}=\dfrac{T_{th}}{k}\\\beta_{th}=\dfrac{T_{th}}{k}\end{cases}
\end{equation}

The thermal fluctuation torque acts on the particle in unit time can be expressed as~\cite{2007_RSI_thermT,li2020NT}

\begin{equation}
\label{Equ_S7}
T_{th}=\sqrt{4k_BT_{emp}I\gamma}
\end{equation}

\noindent
where $k_B$ is the Boltzmann constant, $T_{emp}$ is the ambient temperature, and $\gamma=c/I=1/\tau$ is the rotational friction coefficient of particles, which can be obtained by measuring the decay time $\tau$. The thermal fluctuation torque is equivalent to the input angular velocity $\Omega_{th}$,

\begin{equation}
\label{Equ_S8}
\Omega_{th}=\dfrac{T_{th}}{H}
\end{equation}

Since

\begin{equation}
\label{Equ_S9}
H=I\Omega
\end{equation}

\noindent
and the max rotational frequency $\omega_{max}$ is limited by the ultimate tensile strength $\sigma$=7$GPa$ of the material~\cite{liGHz2018prl,1995_MS_strength}, the max rotational frequency is~\cite{2018_SA_strength}

\begin{equation}
\label{Equ_S10}
\omega_{\mathbf{max}}=\sqrt{\frac{\sigma}{K\rho}}\frac1a
\end{equation}

\noindent
where $K$=0.398 is a shape- and material-dependent constant, $\rho$=2930$kg/m^3$ is the density of vaterite, $a$ is the radius of the spherical rotor.

\begin{figure}[!ht]
\includegraphics[scale=1]{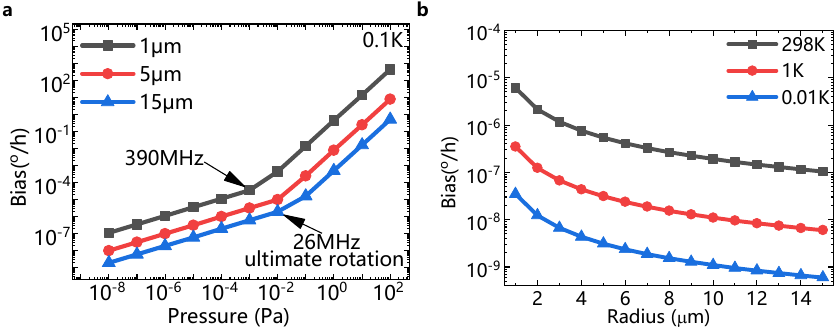}
\caption{\textbf{Thermal noise stability of optically levitated gyroscope} under \textbf{a} different pressure and \textbf{b} particle size in $10^{-8} Pa$.}
\label{fig:S4}
\end{figure}

The bias instability $B$ can be estimated by the equivalent angular velocity

\begin{equation}
\label{Equ_S11}
B=k_\sigma\Omega_{th}=k_\sigma\sqrt{\frac{15k_BK}{8\pi\sigma}}\sqrt{\frac{T_{emp}\gamma}{a^3}}\propto\sqrt{\frac{T_{emp}P_0}{a^3}}
\end{equation}

\noindent
where $k_{\sigma}$=0.304 is the bias instability conversion coefficient of the Standard Deviation Method~\cite{2004_IEEE_Standard}. For the spherical vaterite particles with a diameter of 3.58 $\mu$m used in the experiment, the thermal noise equivalent bias instability is 0.005$^o/s$ when the rotating speed reaches 470$kHz$ at 4$Pa$ without cooling. The measurement results 0.08$^o/s$ is one order of magnitude larger than the calculated value for the disadvantageous reasons, such as center of mass motion that can be decrease by laser cooling, laser power and environment temperature stability that can be suppressed by improving experimental equipment. As shown in Fig. \ref{fig:S4}, the instability of the optically levitated rotor gyroscope can be further improved by decreasing the motion temperature and pressure or increasing the particle size.

\clearpage
\onecolumngrid

\section{Supplementary Figures}
\label{sec:figure}

\begin{figure}[!ht]
\includegraphics[scale=1]{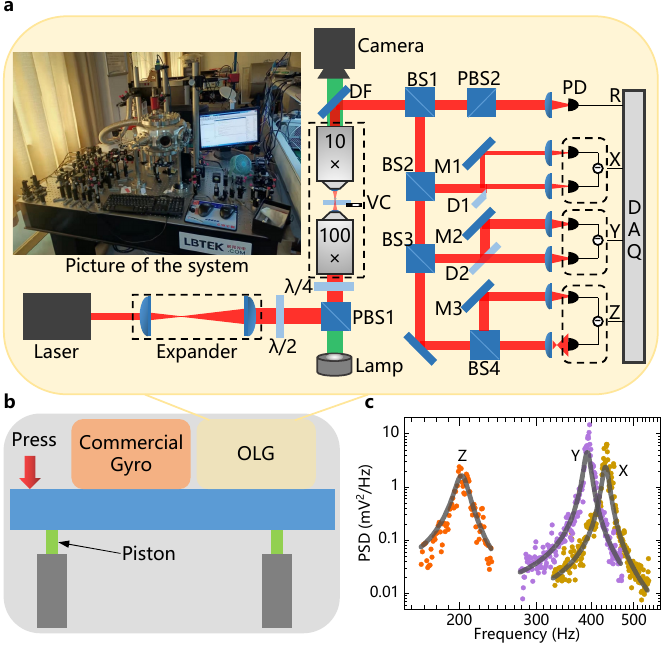}
\caption{\textbf{Experimental setup.} \textbf{a} Labels denote the half-wave plate ($\lambda$/2), polarizing beam splitter (PBS), quarter wave plate ($\lambda$/4), dichroic filters (DF), beam splitter (BS), D-shaped mirror (D), mirror (M), photo detector (PD), vacuum chamber (VC). Inset is the picture of the experimental setup. \textbf{b} Schematic diagram of gyroscopic effect test. The optically levitated rotor gyro (OLG) and the commercial gyro are placed on the air floating platform, and the angular motion of the air floating platform is input by pressing the platform. \textbf{c} Power Spectrum Density (PSD) of the center of mass motion of a vaterite particle trapped by a circularly polarized laser at 40$Pa$}
\label{fig:S5}
\end{figure}

\begin{figure}[!ht]
\includegraphics[scale=1]{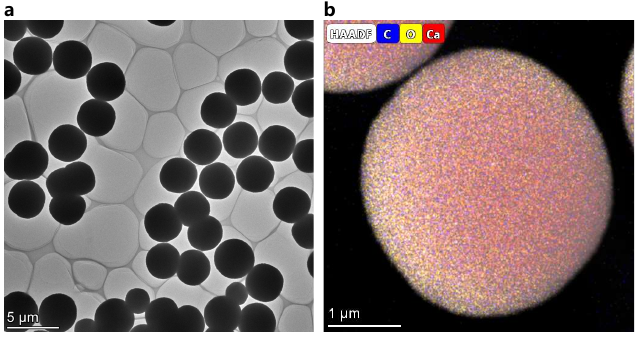}
\caption{\textbf{Shape and element of the vaterite particles.} \textbf{a} Transmission Electron Micro-graph of vaterite particles. \textbf{b} Elemental mapping of one particle.}
\label{fig:S6}
\end{figure}

\begin{figure}[!ht]
\includegraphics[scale=1]{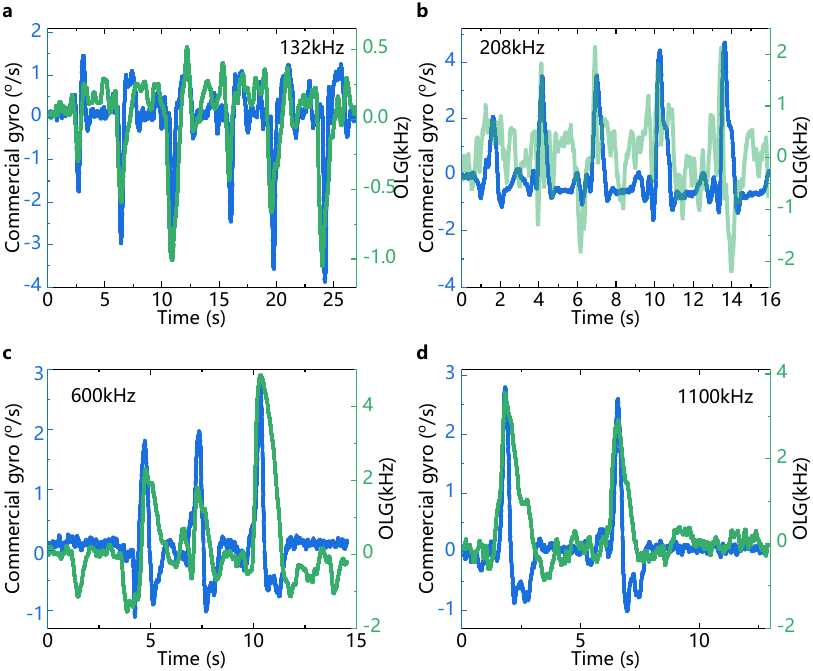}
\caption{Gyroscopic effect of optically levitated particles with different rotational frequency. (OLG: Optically Levitated Gyroscope)}
\label{fig:S7}
\end{figure}

\begin{figure}[!ht]
\includegraphics[scale=1]{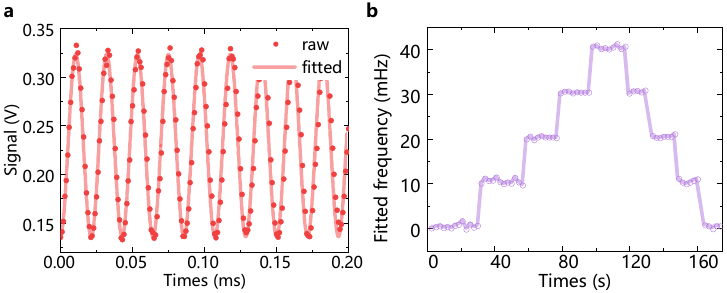}
\caption{\textbf{Measurement method of the instantaneous rotational frequency.} \textbf{a} Raw and fitted data of the rotational signal of a levitated rotor with the rotational frequency 46.7kHz. \textbf{b} The changes of the fitted frequency, and the initial frequency is 100kHz.}
\label{fig:S8}
\end{figure}

\end{document}